\documentclass[cp1251
               ]{jetp} 
\twocolumn 

\begin{document}
{\English

\title{Composite topological objects in topological superfluids}

\setaffiliation1{Low Temperature Laboratory, Aalto University,  \\ P.O. Box 15100, FI-00076 Aalto, Finland}

\setaffiliation2{Landau Institute for Theoretical Physics, \\  acad. Semyonov av., 1a, 142432,
Chernogolovka, Russia}

\setauthor{G.~E.}{Volovik}{12}
\email{volovik@boojum.hut.fi}
 
\rtitle{Landau Institute}
\rauthor{G.~E. Volovik}

 \begin{abstract}
{The spontaneous phase coherent precession of magnetization, discovered in 1984 by Borovik-Romanov, Bunkov, Dmitriev and  Mukharskiy \cite{Borovik1984} in collaboration with Fomin, \cite{HPDtheory},
became now an important experimental tool for study complicated topological objects in superfluid $^3$He.
}
\end{abstract}
\maketitle


\section{Introduction}

Superfluid phases of $^3$He discovered in 1972\cite{OsheroffRichardsonLee1972} opened the new area of the application of topological methods to condensed matter systems. Due to the multi-component order parameter which characterizes the broken $SO(3)\times SO(3)\times U(1)$ symmetry in these phases, there are many inhomogeneous objects -- textures and defects in the order parameter field -- which are protected by topology and are characterized by topological quantum numbers. Among them there are quantized vortices, skyrmions and merons, solitons and vortex sheets, monopoles and boojums, Alice strings, Kibble-Lazarides-Shafi walls terminated by Alice strings, spin vortices with soliton tails, etc.\cite{Volovik2003}   Most of them have been experimentally identified and investigated using nuclear magnetic resonance (NMR) techniquie, and in particular the phase coherent  spin precession discovered in 1984 in $^3$He-B by Borovik-Romanov, Bunkov, Dmitriev and  Mukharskiy \cite{Borovik1984,Borovik1985} in collaboration with Fomin \cite{HPDtheory}.
 Such precessing state, which  has got  the name Homogeneously Precessing Domain (HPD),  is the spontaneously  emerging steady state of precession, which 
preserves the phase coherence across the whole sample even in the absence
of energy pumping and even in an inhomogeneous external magnetic field.
 This spontaneous coherent precession has all the signatures of the coherent superfluid Bose-Einstein condensate of magnons (see review paper \cite{MagnonBECreview}). 

The Bose condensation of magnons in superfluid $^3$He-B had many 
practical applications. 
In Helsinki, owing to the extreme sensitivity  of the Bose
condensate  to textural inhomogeneity, the phenomenon of Bose
condensation  has been applied to studies of  
topological defects by
the HPD spectroscopy.

\section{Superfluid phases of liquid $^3$He}

In bulk liquid $^3$He there are two topologically different superfluid phases, $^3$He-A and $^3$He-B.\cite{VollhardtWoelfle2013}   One is  the chiral superfluid $^3$He-A with topologically protected Weyl points in the quasiparticle spectrum. 
In the ground state of $^3$He-A the order parameter matrix has the form
\begin{equation}
A_{\alpha i}= \Delta_A e^{i\Phi} \hat d_\alpha(\hat e_1^i + i \hat e_2^i)~~,~~ \hat{\bf l}=\hat{\bf e}_1\times \hat{\bf e}_2\,,
\label{Aphase}
\end{equation}
where $\hat{\bf d}$ is the unit vector of the anisotropy in the spin space due to  spontaneous breaking of $SO(3)_S$ symmetry of spin rotations; $\hat{\bf e}_1$ and $\hat{\bf e}_2$ are mutually orthogonal unit vectors;    and $\hat{\bf l}$ is the unit vector of the anisotropy in the orbital space due to  spontaneous breaking of orbital rotations $SO(3)_L$ symmetry. The $\hat{\bf l}$-vector also shows the direction of the orbital angular momentum of the chiral superfluid, which emerges due to spontaneous breaking of time reversal symmetry.  
The chirality of $^3$He-A has been probed in several experiments.\cite{Walmsley2012,Kono2013,Kono2015}

Another phase is the fully gapped  time reversal invariant superfluid $^3$He-B.
In the ground state of $^3$He-B the order parameter matrix has the form
\begin{equation}
A_{\alpha i}= \Delta_B e^{i\Phi} R_{\alpha i}\,,
\label{Bphase}
\end{equation}
where $R_{\alpha i}$ is the real matrix of rotation, $R_{\alpha i}R_{\alpha j}=\delta_{ij}$.
This phase has topologically protected gapless Majorana fermions living on the surface (see reviews \cite{Mizushima2015,Mizushima2016} on the momentum space 
topology in superfluid $^3$He). 

In $^3$He confined in the nematically ordered aerogel (nafen), new phase becomes stable -- the polar phase of $^3$He\cite{Dmitriev2015,HalperinParpiaSauls2018}, whith the order parameter 
\begin{equation}
A_{\alpha i}= \Delta_P e^{i\Phi} \hat d_\alpha\hat m_i\,.
\label{Pphase}
\end{equation}
where orbital vector $\hat{\bf  m}$ is fixed by the nafen strands.
The reason for the appearance of the polar phase in nafen is the analog of the Anderson theorem applied for the polar phase in the presence of the columnar defects (nafen strands), see Refs.
\cite{Fomin2018,Eltsov2019}. While for all the other phases of superfluid $^3$He the transition temperature is suppressed by these impurities.

The polar phase is the time reversal invariant superfluid, which contains Dirac nodal ring in the fermionic spectrum \cite{Volovik2018a,Eltsov2019}.

\section{Strings with solitonic tail}

\subsection{HPD and combined spin-mass vortices with solitonic tail}

\begin{figure}[top]
\centerline{\includegraphics[width=1.0\linewidth]{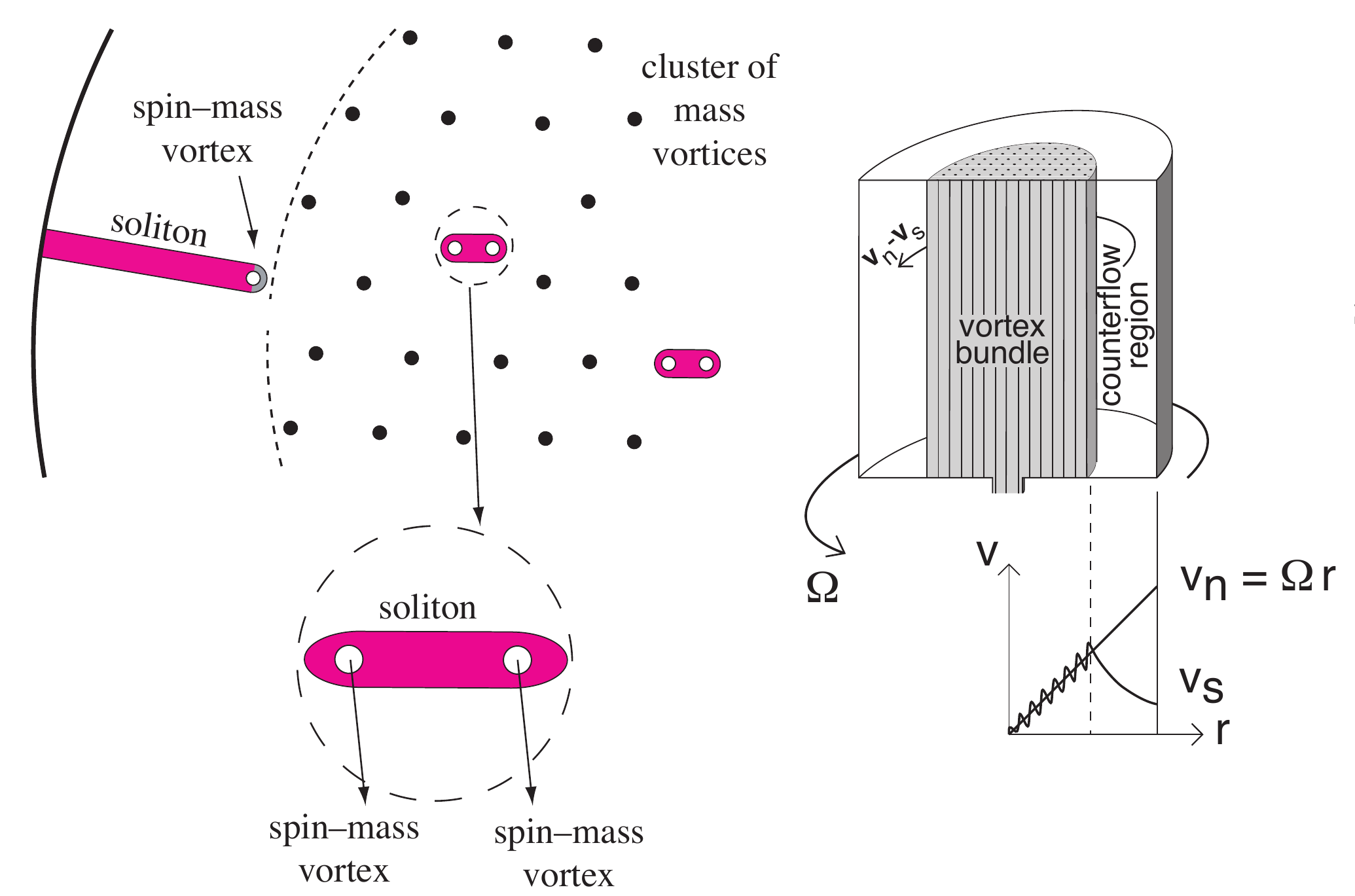}}
\label{SpinMass} 
  \caption{{\it right}: Vortex cluster in rotating container with the vortex free region outside the cluster.
Vortex cluster is formed when starting with the equilibrium vortex state in the rotating container the angular velocity of rotation is increased. The new vortices are not formed if the counterflow in the vortex region does not exceed the critical velocity for vortex formation.
{\it left}: The spin-mass vortex finds its equibrium position on the periphery of the vortex cluster, where the soliton tension is compensated by the Magnus force acting on the mass vortex part of the composite object. The size of the soliton is given by Eq.(\ref{SolitonLength}), and this dependence on the angular velocity of rotation is confirmed by the HPD spectroscopy.
{\it bottom}: The combined object with  ${\cal N}=2$ quanta of circulation: spin-mass vortex + soliton + spin-mass vortex.
}
\end{figure}

There are different types of the topological defects in the  $^3$He-B. Among them 
there are the conventional mass vortices with the ${\cal N}$ winding number of the phase $\Phi$, and the $Z_2$ spin vortex -- the nontrivial winding of the matrix $R_{\alpha i}$. Due to spin-orbit coupling the spin vortex serves as the termination line of the topological soliton wall. Because of the soliton tension the spin vortex moves to the wall of the vessel and escapes the observation.
However, the help comes from the mass vortices. The mass and spin vortices are formed by different fields. They do not interact since they "live in different
worlds". The only instance, where the spin and mass vortices interact,
arises when the cores of a spin and a mass vortex happen to get close to each other and it becomes energetically
preferable for them to form a common core. Thus by trapping the spin vortex on a mass vortex the combined
core energy is reduced and a composite object $Z_2$-string + soliton + mass vortex, or spin-mass vortex
is formed. This object is stabilized near the edge of the vortex cluster in the rotating cryostat, see Fig. 1. 

These combined objects 
have been observed and studied using  HPD spectroscopy \cite{Kondo1992,Korhonen1993}.
  The additional absorption observed in the homogeneously precessing domain (HPD)  is proportional to the soliton area $A=lh$, where $h$ is the heght of the container, and $l$
 is the length of the cross-section of the soliton. In the rotating container the length $l$ is given by the width of the counterflow vortex-free zone, which is regulated by changing the angular velocity of rotation $\Omega$ at fixed number $N$ of vortices in the cell:
\begin{equation}
l(\Omega)= R\left(1- \sqrt{\frac{\Omega_V(N)}{\Omega}} \right) \,,
\label{SolitonLength}
\end{equation}
Here $R$ is the radius of the cylindrical container, and $\Omega_V(N)$ is the angular velocity in the state in the rotating container with  equilibrium number of vortices $N=2\pi R^2\Omega_V(N)/\kappa$.
The equilibtium state is obtained by cooling through $T_c$ under rotation, and then we increase the angular velocity of rotation,  $\Omega>\Omega_V(N)$. The new vortices are not created because of high energy barrier, and as a result the counterflow region appears. 
The dependence of the attenuation of the HPD state follows Eq.(\ref{SolitonLength}) \cite{Kondo1992,Korhonen1993}.

\subsection{Alice strings with and without solitonic tail}

The half-quantum vortices (HQVs) were originally suggested to exist in the chiral superfluid $^3$He-A \cite{VolovikMineev1976}. 
The half-quantum vortex represents the condensed matter analog of the Alice string in particle physics.\cite{Schwarz1982}  The HQV is the vortex with fractional circulation of superfluid velocity, ${\cal N}=1/2$. It is topologically confined with the fractional spin vortex, in which $\hat{\bf d}$ changes sign when circling around the vortex:
\begin{equation}
\hat{\bf d}(\hat{\bf r}) e^{i\Phi(\hat{\bf r})}=\left(\hat{\bf x} \cos\frac{\phi}{2}+ \hat{\bf y} \sin\frac{\phi}{2}\right)e^{i\phi}
\,,
\label{HalfQuantumVortex}
\end{equation}
When the azimuthal coordinate $\phi$ changes from 0 to $2\pi$ along the circle around this object, the vector $\hat{\bf d}(\hat{\bf r})$  changes sign and simultaneously the phase $\Phi$ changes by $\pi$, giving rise to ${\cal N}=1/2$. The  order parameter (\ref{HalfQuantumVortex}) remains continuous along the circle.
While a particle that moves around an Alice string flips its charge, the quasiparticle moving around the half-quantum vortex flips its spin quantum number. This gives rise to the Aharnov-Bohm effect for spin waves
in NMR experiments.\cite{SalomaaVolovik1987}

\begin{figure}[top]
\centerline{\includegraphics[width=1.0\linewidth]{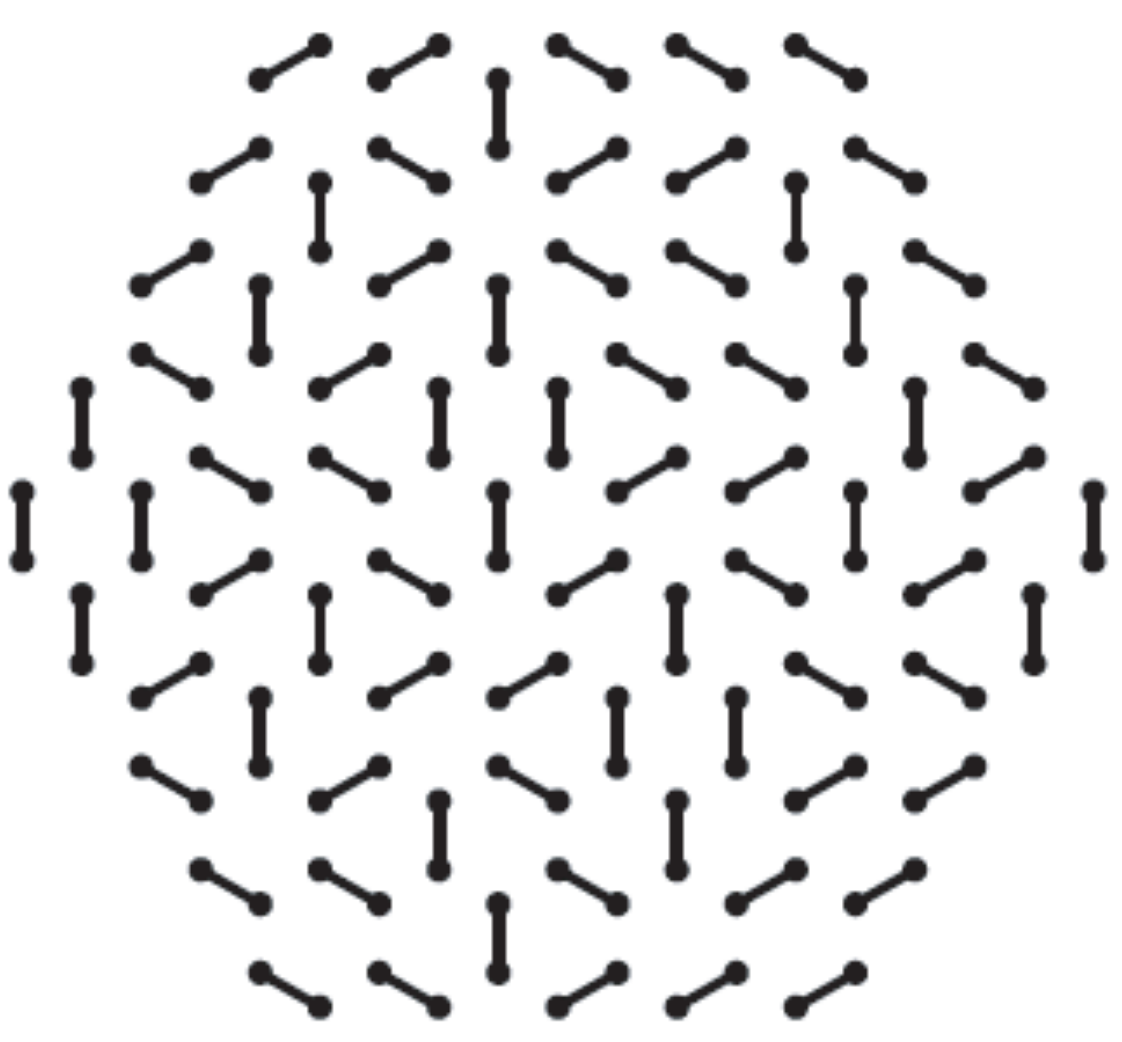}}
\label{SolitonLatticeFig} 
  \caption{Illustration of the lattice of solitons emerging between the Alice strings (half-quantum vortices) in the polar phase if $^3$He, when the magnetic field is tilted with respect of the aerogel strands. The half quantum vortices survive the soliton tension because they are pinned by the strands. The NMR measurements give information on the total length of the soliton and thus on the number of the Alice strings in the cell.
}
\end{figure}

However, before being experimentally observed in $^3$He-A, the HQVs were first observed in another topological phase of $^3$He -- the polar phase \cite{Autti2016}. The reason for that is that in $^3$He-A the spin-orbit interaction chooses the preferrable orientation for the vector $\hat{\bf d}$ describing the spin degrees of freedom of the order parameter. This leads to formation of a soliton interpolating between two degenerate vacua with $\hat{\bf d}=\hat{\bf l}$ and $\hat{\bf d}=-\hat{\bf l}$.
 The energy of soliton prevents the nucleation of the Alice strings in $^3$He-A.

 In contrast, in the polar phase the spin-orbit interaction can be controlled not prohibit the formation of HQVs.  In the absence of magnetic field, or if the field is along the nafen strands the
spin-orbit interaction does not lead to formation of the solitons attached to the spin vortices. As a result the half-quantum vortices become emergetically favorable and appear in the rotating cryostat if the sample is cooled down from the normal state under rotation.

Nevertheless the solitons help to observe the Alice string first in polar phase and after that in the polar distorted A-phase. 
In the polar phase, when the orientation of the magnetic field is tilted with respect to aerogel strands, the spin-orbit interaction generates the solitons between the half-quantum vortices. But the Alice strings are strongly pinned by the nafen strands, and the soliton cannot shrink, see 
Fig. 2. The HQVs are identified due to peculiar dependence of the NMR frequency shift on the tilting angle of magnetic field \cite{Autti2016}.
The NMR experiments also allow measure the density of the Alicie string by measuring the soliton density.

Due to the strong pinning, the Alice strings formed in the polar phase by rotation of the superfluid or by the Kibble-Zurek mechanism, survive the transition to the $^3$He-A (actually to the distorted A-phase) \cite{Makinen2019}.

\section{HPD and KLS wall between Alice strings}

\subsection{Nonaxisymmetric vortex in $^3$He-B as KLS wall bounded by Alice strings}

The mass vortices in $^3$He-B  are presented in several  forms. In particular, a pair of spin-mass vortices may form a molecule, where the soliton serves as chemical bond.  As a result one obtains the doubly quantized vortex, i.e. with ${\cal N}=2$ circulation quanta, see 
Fig. 1. Such vortex molecules have been also identified in HPD spectroscopy \cite{Kondo1992,Korhonen1993}.

The "conventional" ${\cal N}=1$ vortex has also an unusual structure in $^3$He-B. Already in the first experiments with rotating 
 $^3$He-B the first order phase transition has been observed, which has been associated with the transition inside the vortex core \cite{Ikkala1982}.
It was suggested that at the transition the vortex core becomes 
non-axisymmetric, i.e. the axial symmetry of the vortex is spontaneously broken in the vortex core\cite{Thuneberg1986,VolovikSalomaa1985}. This was confirmed in the further experiments where the coherently precessing magnetization was used \cite{Kondo1991}.

In the weak coupling BCS theory, which is applicable at low pressure, such vortex can be considered as splitted into two half-quantum vortices connected by the domain  wall \cite{Volovik1990,SilaevThuneberg2015}, which is the analog of the Kibble-Lazarides-Shafi wall bounded by cosmic strings
\cite{Kibble1982}.
 The separation between the half-quantum vortices increases with decreasing
pressure.

The phenomenon of the additional symmetry breaking in the core of the topological defect has been also discussed for cosmic strings\cite{Witten1985}. The spontaneous breaking of the electromagnetic $U(1)$ symmetry in the core of the cosmic string has been considered, due to which the core becomes superconducting.

\begin{figure}[top]
\centerline{\includegraphics[width=1.0\linewidth]{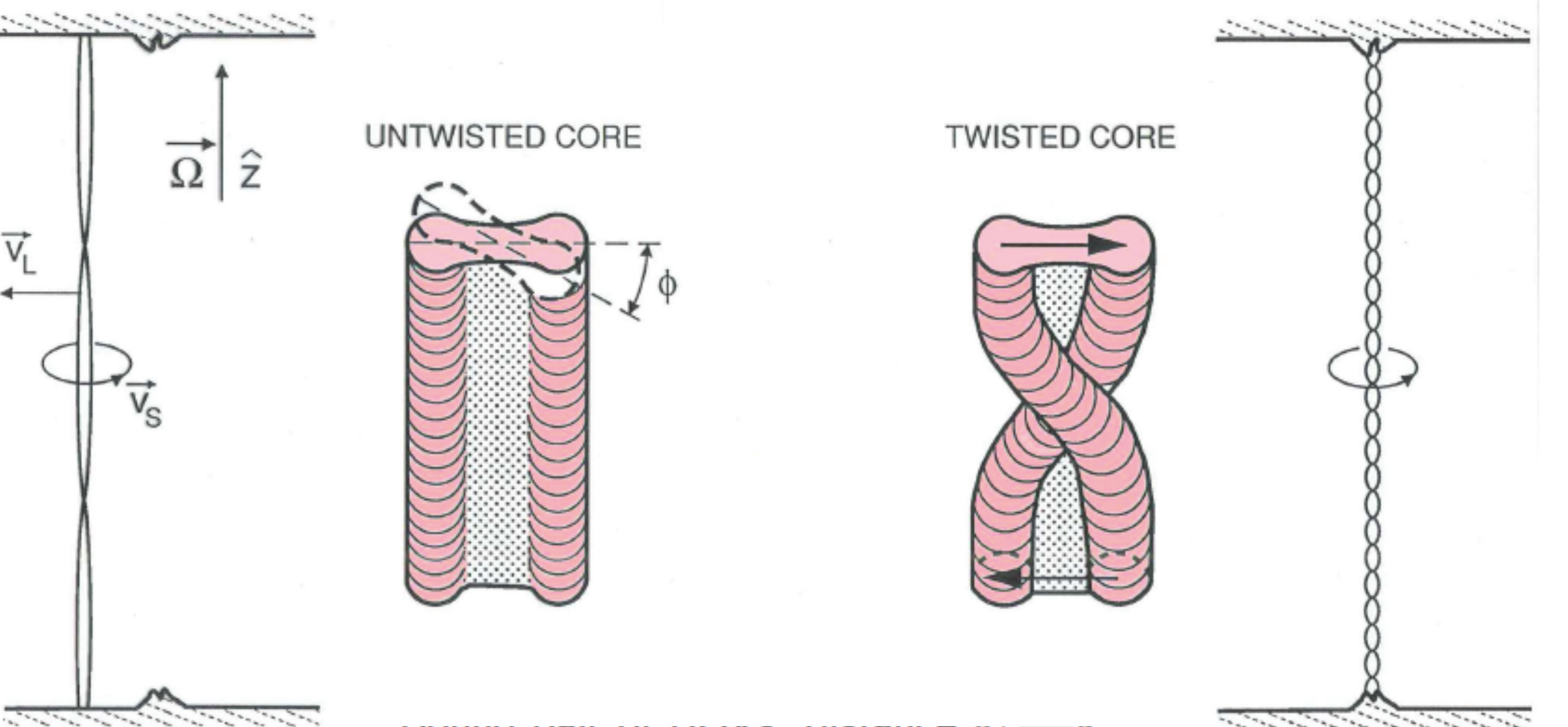}}
\label{VortexTwistfig} 
  \caption{The vortex in $^3$He-B with the non-axisymmetric core as a pair of Alice strings connected by Kibble-Lazarides-Shafi wall. The HPD with its coherent precession of magnetization is used to twist the core. The vortex with twisted core is analogous to Witten superconducting string with the electric current along the string core \cite{Witten1985}.
}
\end{figure}

For the $^3$He-B vortices, the spontaneous breaking of
the
$SO(2)$ symmetry in the
core leads to the Goldstone bosons -- the modes in which the degeneracy
parameter, the axis of anisotropy ${\bf b}$ of the vortex core, is
oscillating. The homogeneous magnon condensate, the HPD state, has been used to study  the structure and twisting dynamics of this non-axisymmetric core. The coherent precession of magnetization excites the vibrational Goldstone mode via spin-orbit interaction. Moreover, due to spin-orbit interaction the precessing magnetization rotates the core around its axis with constant
angular velocity. In addition, since the core was pinned on the top and
the bottom of the container, it was possible even to screw the core (see Figs. 3 and 4).
 Such a twisted core
corresponds to the Witten superconducting string with the electric supercurrent
along the core. The rigidity of twisted core differs from that of the straight core, which is clearly seen in HPD experiments, see 
Fig. 4.

Oscillations of the vortex core under coherent spin precession also lead to the observed radiation of acoustic magnon modes \cite{Zavjalov2016}.

\begin{figure}[top]
\centerline{\includegraphics[width=1.0\linewidth]{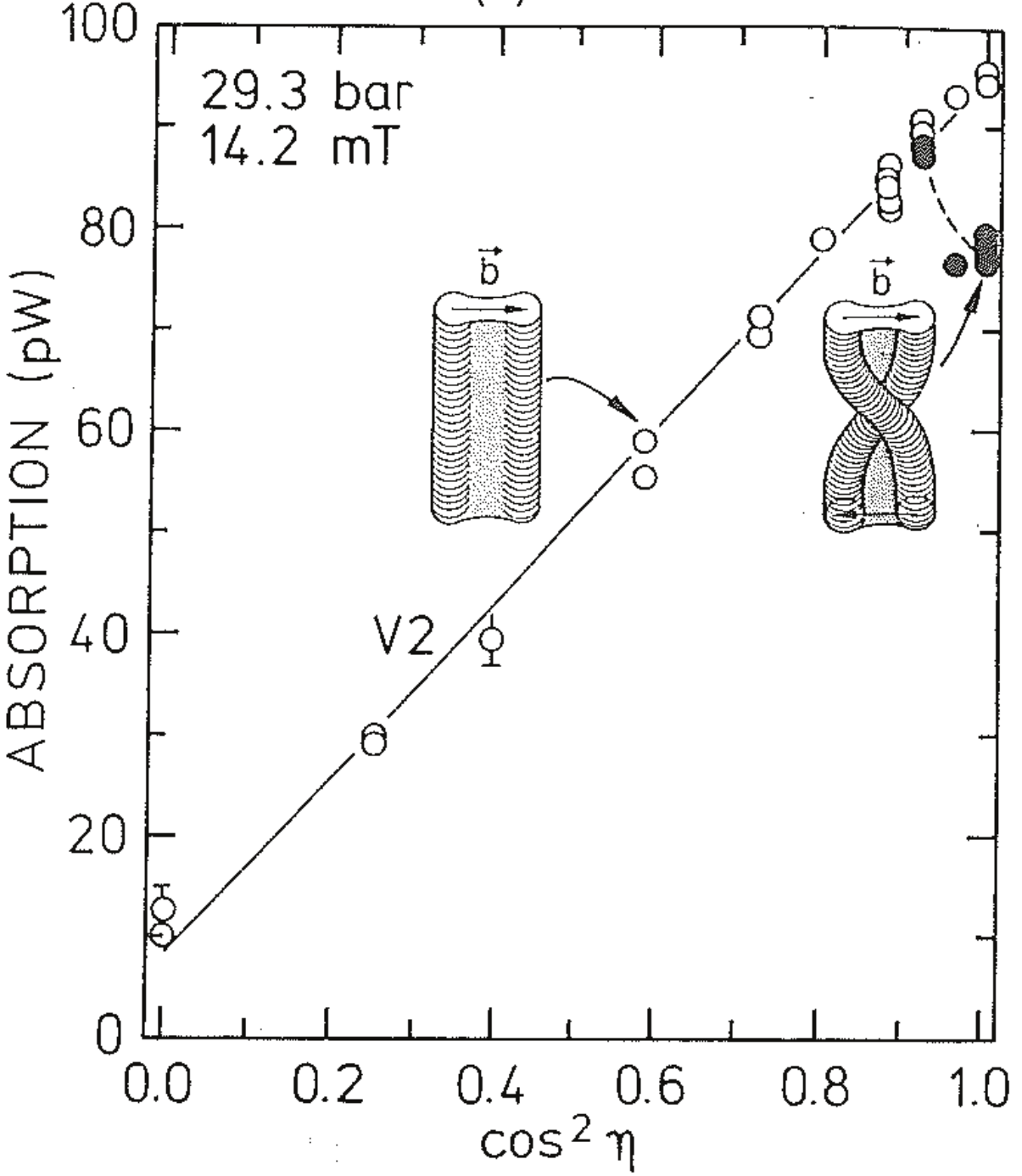}}
\label{VortexTwist2fig} 
  \caption{HPD absorbtion as the function of the tilting angle $\eta$ of magnetic field in case of the  Witten strings with twisted core (filled circles) and strings with untwisted core (open circles). The estimated critical angle at which the tilted field prevents twisting by HPD is in agreement with experiment.
}
\end{figure}

\subsection{Alice strings with KLS wall in polar distorted B-phase}

\begin{figure}[top]
\centerline{\includegraphics[width=1.0\linewidth]{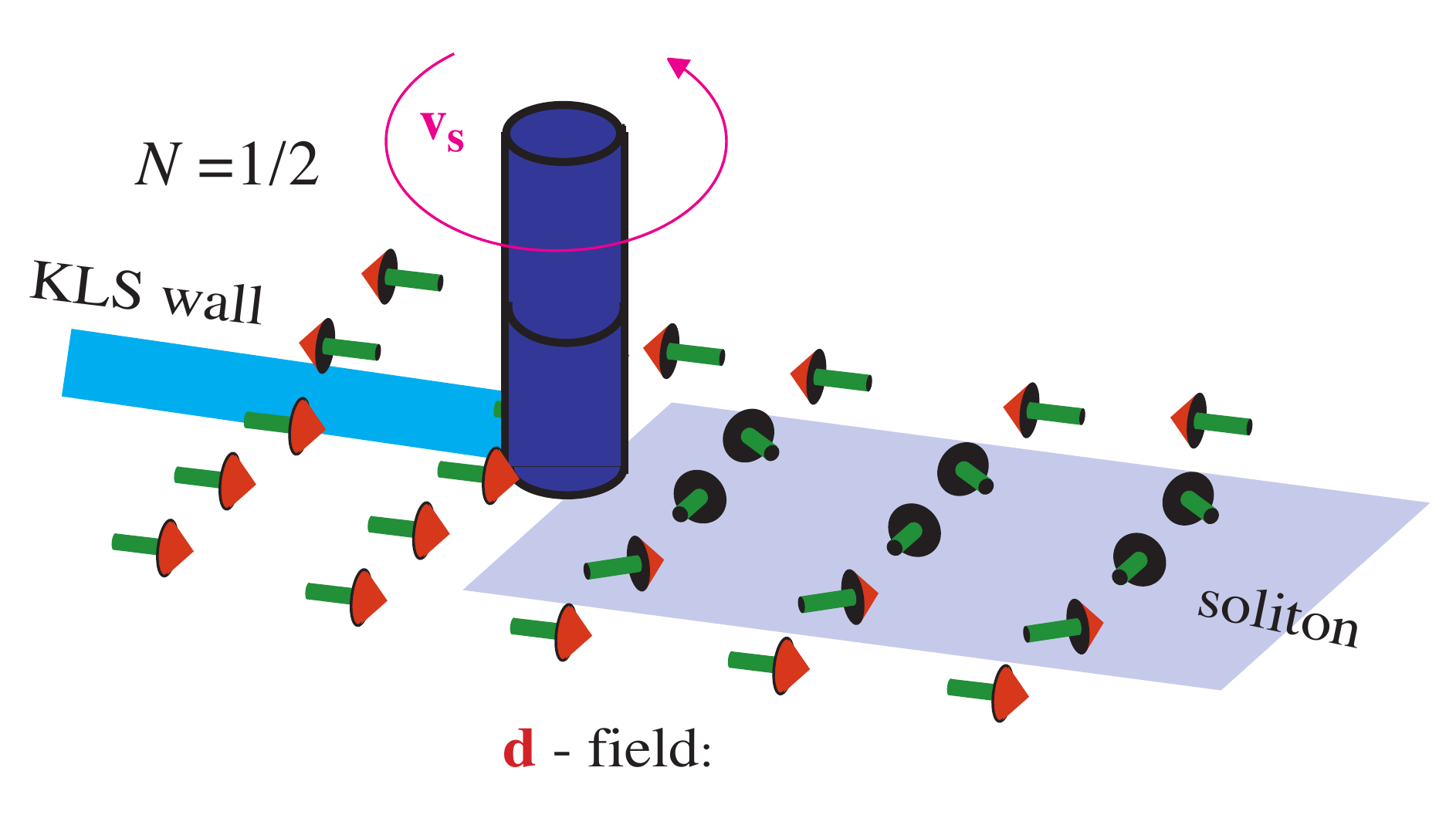}}
\label{KLSfig} 
  \caption{The Alice string terminating the Kibble-Lazarides-Shafi wall in the polar distorted B-phase in nematical aerogel.
Due to the pinning of Alice string by the areogel strands the KLS wall can be arbitrarily long: the wall tension is unable to unpin the string. In addition to the KLS wall there is also the soliton tail of the string. As a result one has  the triple object: KLS wall + Alice string + soliton.
}
\end{figure}

In the  vortices with asymmetric cores the equilibrium distance between the Alice strings is rather small. 
The essentially larger KLS walls between the strings have been observed in the B-phase in nafen \cite{Makinen2019}, see Fig. 5.
It appeared that the Alice strings formed in the polar phase by rotation of the superfluid or by the Kibble-Zurek mechanism, survive the transition to the $^3$He-B (actually to the distorted B-phase). They remain pinned, in spite of the formation of the KLS walls between them.

This allows us to study the unque properties of the KLS wall. In particular, the KLS wall separates two degenerate cvacua with different signs of the tetrad determinant, and thus between the "spacetime" and "antispacetime" \cite{Volovik2019}.

\section{Combined objects to be observed}

\subsection{Multi-quantum vortex as closed vortex sheet}

\begin{figure}[top]
\centerline{\includegraphics[width=1.0\linewidth]{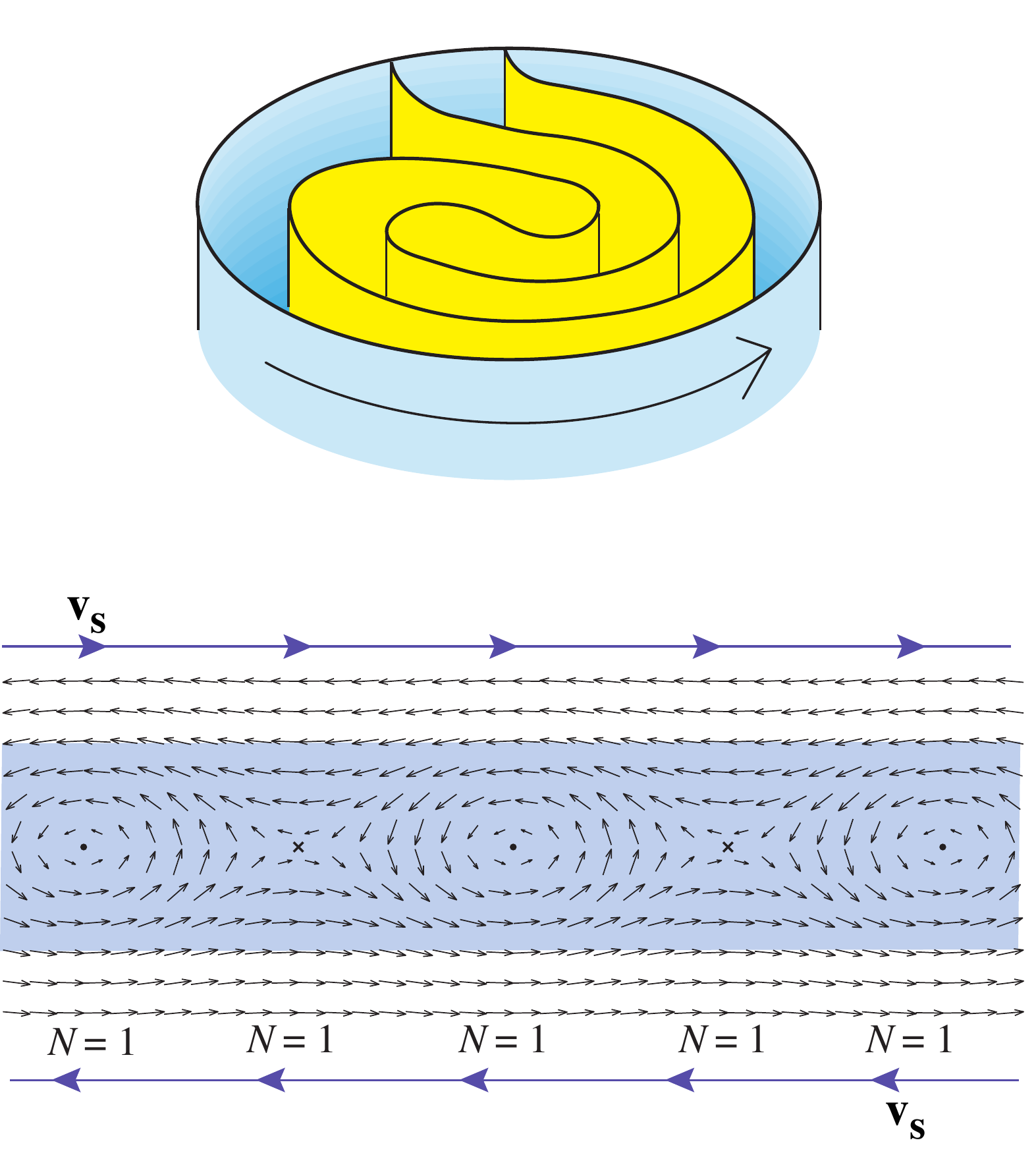}}
\label{VortexSheetfig} 
  \caption{{\it top}: Typical vortex sheet in $^3$He-A in rotating container. It mimics the system of the equidistant  cylindrical vortex sheets suggested by Landau and Lifshitz for the descrption of the rotating superfluid \cite{LandauLifshitz1955}.
{\it bottom}: The element of the vortex sheet in $^3$He-A. The vortex sheet is the soliton, which contains kinks in terms of merons.  Each meron has circulation quantum $N=1$. There are different scenarios in which the vortex sheets with different geometries are prepared in the experiments (see Ref.\cite{EltsovKopnin2002}).
}
\end{figure}

In the chiral superfluid, the superfluid velocity $ {\bf v}_{\rm s}$ of the chiral condensate is determined not only by the condensate phase $\Phi$, but also by  the  orbital triad  $\hat{\bf e}_1$, $\hat{\bf e}_2$ and $\hat{\bf l}$:
\begin{equation}
  {\bf v}_{\rm s}=\frac{\hbar}{2m}\left(\nabla\Phi+\hat{\bf e}_1^i\nabla \hat{\bf e}_2^i \right)\,,
\label{SuperfluidVelocity}
\end{equation}
where $m$ is the mass of the $^3$He atom. 
As distinct form the non-chiral superfluids, where the vorticity is presented in terms of the quantized singular vortices with the phase winding $\Delta\Phi = 2\pi {\cal N}$ around the vortex core, in $^3$He-A the vorticity can be continuous. The continuous vorticity is represented by the texture of the unit vector $\hat{\bf l}$ according to the Mermi-Ho relation:\cite{Mermin-Ho}
\begin{equation}
  \nabla\times{\bf v}_{\rm s} =
     \frac{\hbar}{4m}e_{ijk} \hat l_i{\bf \nabla} \hat l_j\times{\bf
\nabla}
\hat l_k \,.
\label{Mermin-HoEq}
\end{equation}
Experimentally the continuous vorticity is typically observed in terms of skyrmions (or
the Anderson-Toulouse-Chechetkin vortices \cite{Chechetkin1976,AndersonToulouse1977}), see the upper part of Fig. 8.
Each skyrmion has ${\cal N}=2$ quanta of circulation of superfluid velocity. The skyrmion can be also presented as the combination of two merons with ${\cal N}=1$ each.

\begin{figure}[top]
\centerline{\includegraphics[width=1.0\linewidth]{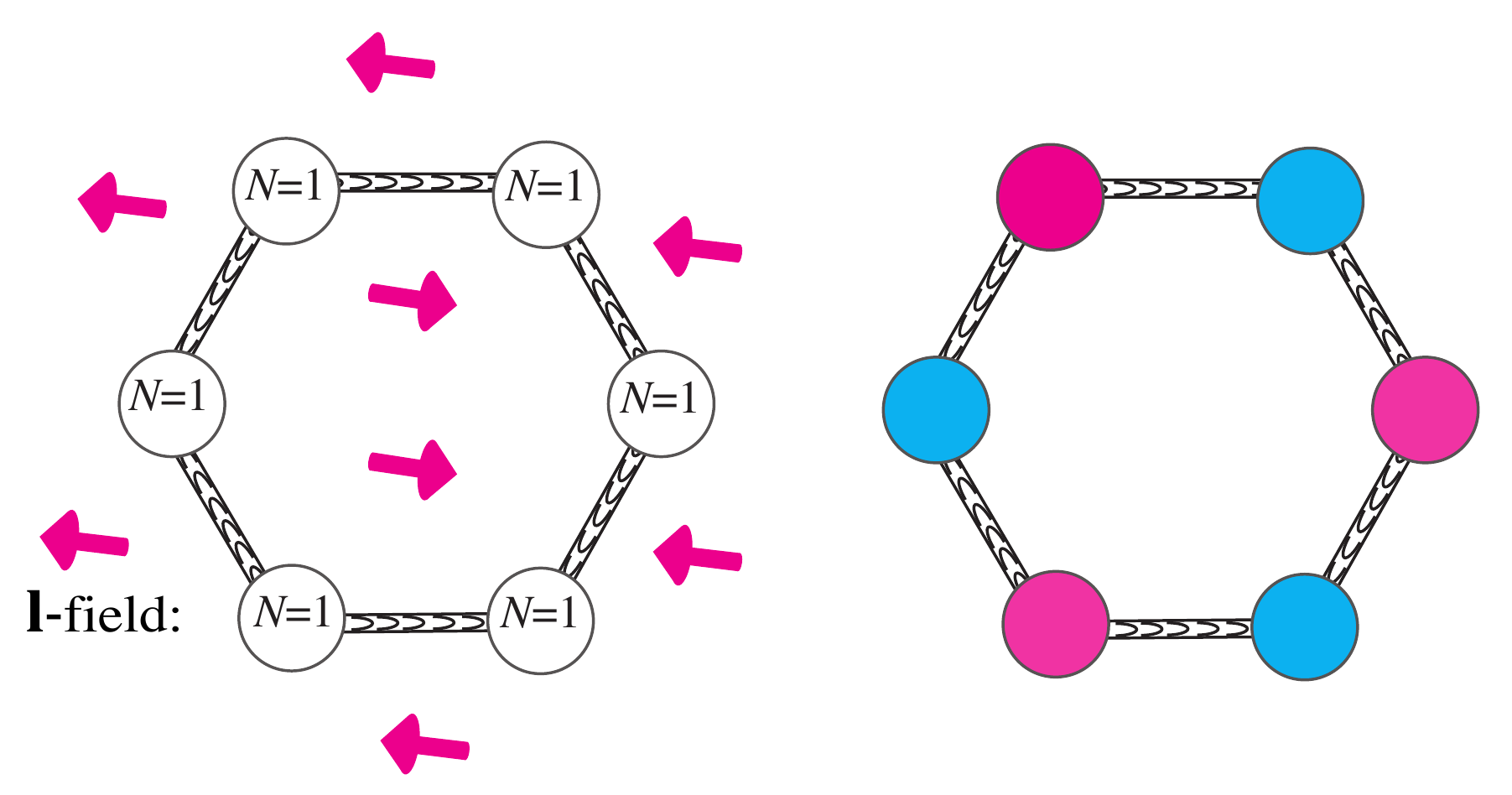}}
\label{MultiQuantafig} 
  \caption{{\it left}: The multi-quantized vortex can be stabilized as the closed vortex sheet: cylindrical soliton with merons \cite{KrusiusVolovik1994}.
The tension of the soliton is compensated by repulsion of vortices (merons). {\it right}: The cosmic analog of this composite object: cosmic necklace\cite{Shafi2019}. Monopoles and/or antimonopoles are joined together by flux tubes.   
}
\end{figure}

In 1994  a new type of continuous vorticity has been observed  in $^3$He-A -- the vortex texture in the form of the vortex sheets \cite{Parts1994a,Parts1994b,EltsovKopnin2002}, see Fig. 6 {\it top} with a single vortex sheet in container.  Vortex sheet is the topological soliton with kinks, each kink representing  the continuous Mermin-Ho vortex with ${\cal N}=1$ circulation of superfluid velocity, 
Fig. 6 {\it bottom}.

In principle, using the vortex sheet one may construct the continuous vortices with arbitrary even number ${\cal N}=2k$ circulation quanta. This is the   soliton forming the
closed cylindrical surface, which contains ${\cal N}$ "quarks"  -- merons, \cite{KrusiusVolovik1994,VolovikVortexSheet2015}  see
Fig. 7  {\it left} for ${\cal N}=6$. However, such multi-quantum vortices are still waiting for their  observation.

\subsection{Monopoles, necklaces and monopole lattices with Alice strings}

\begin{figure}[top]
\centerline{\includegraphics[width=1.0\linewidth]{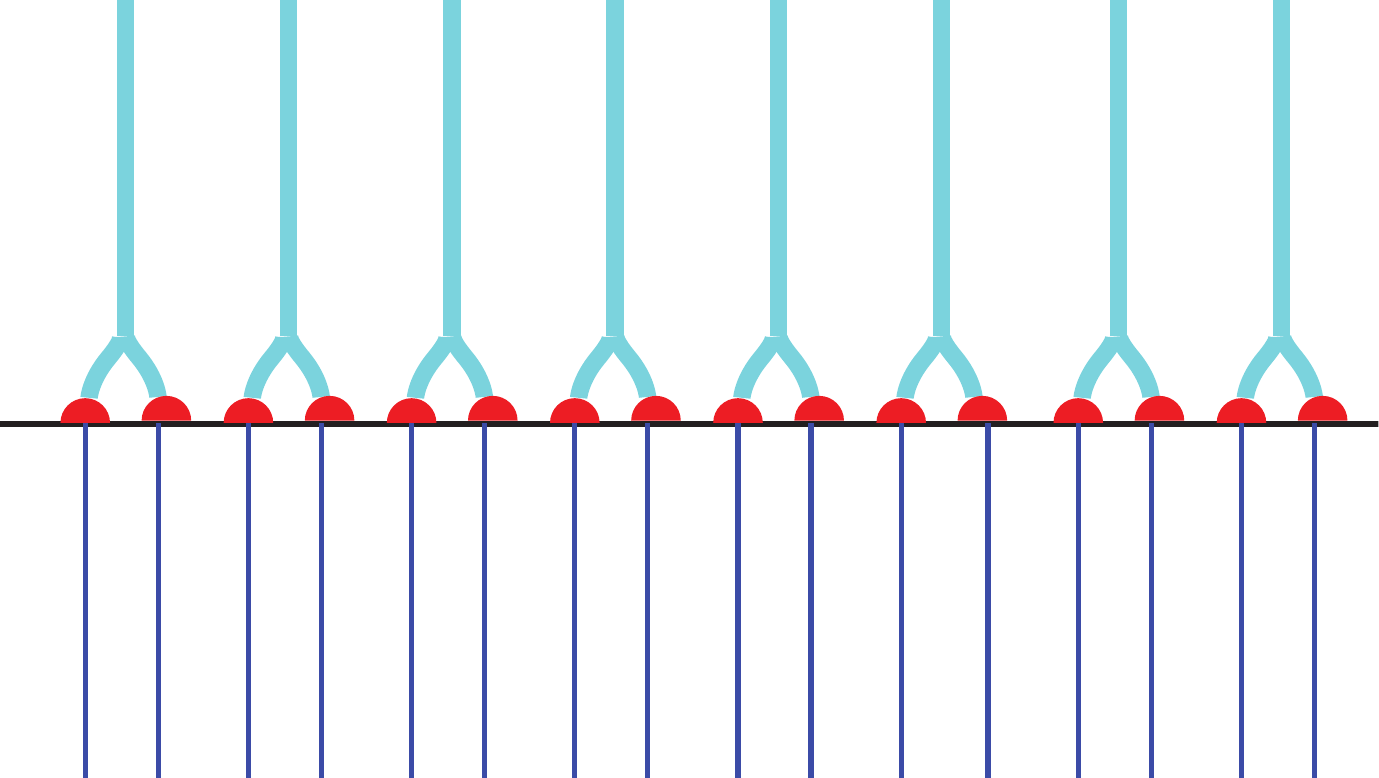}}
\label{Boojumsfig} 
  \caption{Skyrmion in the A-phase splits into two merons. Each meron is terminated by boojum --  the point topological objects, which lives at the interface between A-phase and B-phase. Boojum also plays the role of the Nambu monopole, which terminates the string - the $N=1$ vortex on the B-side of the interface. 
}
\end{figure}

Another object which is waiting for its observation in $^3$He-A is the vortex terminated by hedgehog\cite{Blaha1976,VolovikMineev1976a}. This is the condensed matter analog of the electroweak magnetic monopole and the other monopoles connected by strings \cite{Kibble2008}. The hedgehog-monopole, which terminates the vortex, exists in particular at the interface between $^3$He-A and $^3$He-B. The topological defects living on the surface of the condensed matter system or at the interfaces are called boojums
\cite{Mermin1977}. They are classified in terms of relative homotopy groups \cite{Volovik1978}.   Boojums terminate the 
$^3$He-B vortex-strings with ${\cal N}=1$. The boojums do certainly exist on the surface of rotating $^3$He-A and at the interface between the rotating $^3$He-A and $^3$He-B \cite{Krusius2003}, see Fig. 8. However,  at the moment their NMR signatures are too weak to be resolved in NMR experiments in $^3$He. But the vortex terminated by the hedgehog-monopole was observed in cold gases \cite{Mottonen2014}.

\begin{figure}[top]
\centerline{\includegraphics[width=1.0\linewidth]{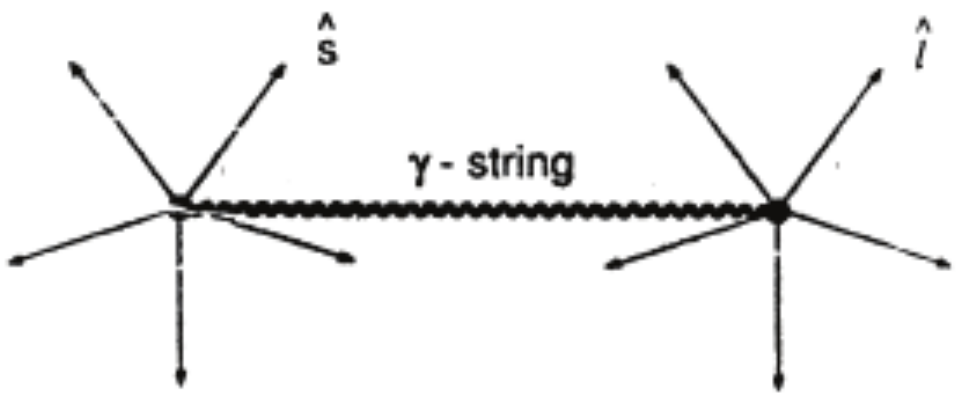}}
\label{HPDhedgehogfig} 
  \caption{ Spin and orbital hedgehogs in magnon BEC (HPD) connected by string from Ref.\cite{MisirpashaevVolovik1992}. 
}
\end{figure}

The HPD state has its own topological defects \cite{MisirpashaevVolovik1992}, and among them are the spin and orbital monopoles connected by string in Fig. 9.

In particle physics the monopoles terminating strings are called Nambu monopoles \cite{Nambu1977}.
Several monopoles connected by strings may form the multi-monopole objects, such as necklace in 
Fig. 7 ({\it right})\cite{Shafi2019}. 
This is similar to the vortex sheet necklace in 
Fig. 7 ({\it left}).

\begin{figure}[top]
\centerline{\includegraphics[width=1.0\linewidth]{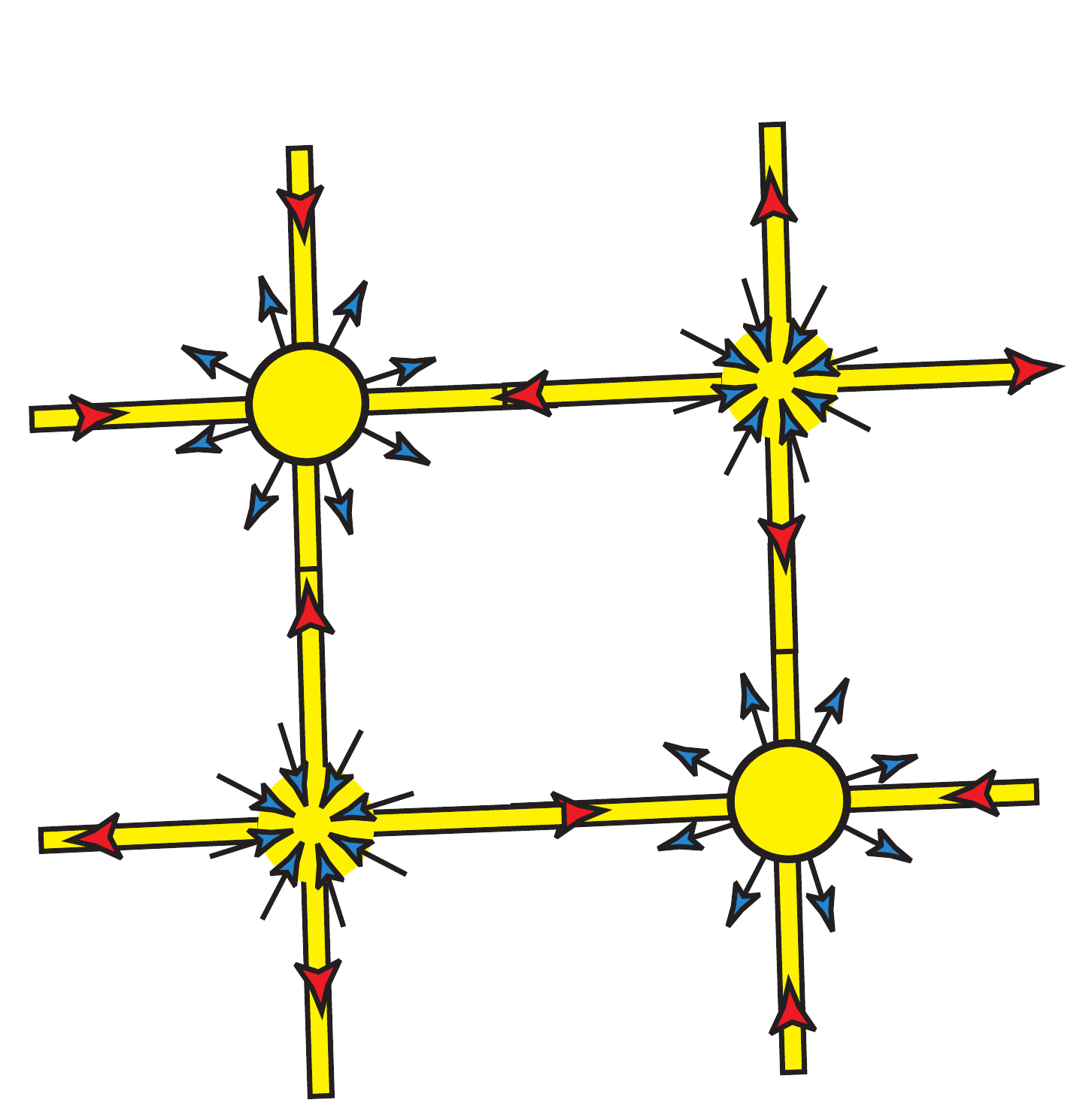}}
\label{2Dfig} 
  \caption{Two dimensional lattice of monopoles (hedgehgogs in the $\hat{\bf l}$-field) joined together by Alice strings (half-quantum vortices). Each monopole is the source or sink of 4 strings. 
}
\end{figure}

\begin{figure}[top]
\centerline{\includegraphics[width=1.0\linewidth]{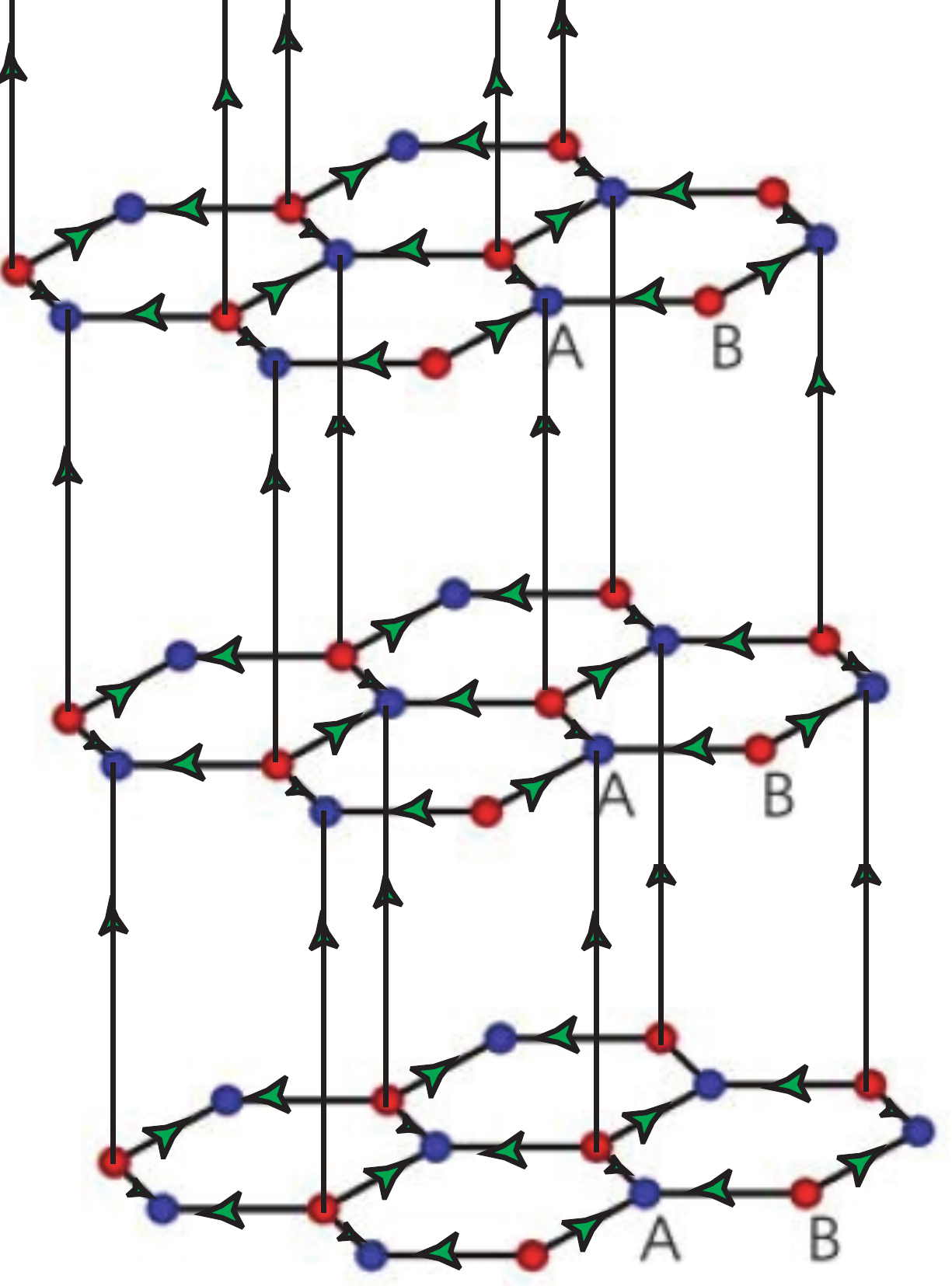}}
\label{3Dfig} 
  \caption{Three dimensional lattice of monopoles (on sites A) and anti-monopoles (on sites B), which are joined together by Alice strings (half-quantum vortices).
}
\end{figure}

In $^3$He-A the analogs of Nambu monopoles and Alice strings may form the more complex combinations.
This is because the monopole serves as a source or sink of ${\cal N}=2$ circulation quanta, and thus can be the termination point of 4 Alice strings with ${\cal N}=1/2$ each.
This in particular allows construct the 2D and 3D lattices of monopoles, in 
Fig. 10 and in
Fig. 11 correspondingly.

\section{Conclusion}

Here we considered several  types of  the topological confinement. The composite topological objects were experimentally observed in superfluid $^3$He by using the unique phenomenon of HPD -- the spontaneously formed coherent precession of magnetization discovered by the Borovik-Romanov group in Kapitza Institute. With HPD spectroscopy, two key  objects have been identified in $^3$He-B: spin-mass vortex  [$Z_2$ spin vortex+ soliton+mass vortex] and non-axisymmetric vortex 
[Alice string + Kibble-Lazarides-Shafi wall + Alice string]. 
One may expect the other more complicated examples of the topological confinement  of the objects of different dimensions. The complicated composite objects, such as nexus, live also in the momentum space of topological materials \cite{TeroNexus2015}.

{\bf Acknowledgements}. This work has been supported by the European Research Council (ERC) under the European Union's Horizon 2020 research and innovation programme (Grant Agreement No. 694248).
I thank Q. Shafi for the discussions that ultimately led to this article.

\end{document}